\documentclass{aa}
\usepackage{natbib}
\usepackage{graphicx}
\usepackage{txfonts}
\usepackage{adjustbox}

\newcommand{\de}{\ensuremath{\mathrm{d}}}

\newcommand{\ob}{^{\rm ob}}
\newcommand{\true}{^{\rm true}}

\def\bm {\bf}

\begin{document}
   \title{CODEX clusters.}
   \subtitle{ The Survey, the Catalog, and Cosmology of the X-ray Luminosity Function.}

   \author{A. Finoguenov \inst{1}
	\and E. Rykoff \inst{2}
        \and N. Clerc \inst{3}
        \and M. Costanzi \inst{4}
        \and S. Hagstotz \inst{5}
        \and J. Ider Chitham\inst{6}
        \and K. Kiiveri \inst{1,13}
        \and C. C. Kirkpatrick \inst{1,13}
        \and R. Capasso \inst{5}
        \and J. Comparat\inst{6}
        \and S. Damsted\inst{1}
        \and R. Dupke \inst{7}
        \and G. Erfanianfar\inst{6} 
        \and J. Patrick Henry\inst{8}
        \and F. Kaefer \inst{6}
        \and J-P. Kneib\inst{9}
        \and V. Lindholm \inst{1,13}
        \and E. Rozo \inst{10}
        \and L. van Waerbeke\inst{11}
	\and J. Weller\inst{12,6}
    }         

   \institute{
Department of Physics, University of Helsinki, Gustaf H\"allstr\"omin
katu 2, FI-00014 Helsinki, Finland
\and
Kavli Institute for Particle Astrophysics \& Cosmology, PO Box 2450, Stanford University, Stanford, CA 94305, USA; \\ \; SLAC National Accelerator Laboratory, Menlo Park, CA 94025, USA
\and
IRAP, Université de Toulouse, CNRS, UPS, CNES, Toulouse, France
\and
INAF-Osservatorio Astronomico di Trieste, Via G. B Tiepolo 11, I-34143 Trieste, Italy; \\ \; IFPU-Institute for Fundamental Physics of the Universe, Via Beirut 2, 34014 Trieste, Italy
\and
Oskar Klein Centre, Department of Physics, Stockholm University, AlbaNova University Centre, SE 106 91 Stockholm, Sweden
\and
Max Planck Institute for Extraterrestrial Physics, Giessenbachstrasse, D-85748 Garching, Germany
\and
Observatório Nacional, Ministério da Ciência, Tecnologia, Inovação e Comunicações, Rua General José Cristino, 77, São Cristóvão, 20921-400 Rio de Janeiro, RJ, Brazil
\and
Institute for Astronomy, 2680 Woodlawn Drive, Honolulu, HI 96822, USA
\and
Laboratoire d'Astrophysique, Ecole Polytechnique F\'ed\'erale de Lausanne (EPFL), Observatoire de Sauverny, CH-1290 Versoix, Switzerland
\\ \;
Aix Marseille Universit\'e, CNRS, LAM (Laboratoire d'Astrophysique de Marseille) UMR 7326, 13388, Marseille, France
\and
Department of Physics, University of Arizona, Tucson, AZ 85721, USA
\and
Department of Physics and Astronomy, University of British Columbia, 6224 Agricultural road, Vancouver, BC V6T 1Z1, Canada
\and
Universit\"{a}tssternwarte M\"{u}nchen, Scheinerstrasse 1, 81679 M\"{u}nchen, Germany \\ \;
Excellence Cluster Origins, Boltzmannstra\ss e 2, D-85748 Garching, Germany
\and
Helsinki Institute of Physics, Gustaf H\" allstr\" omin katu 2, University of Helsinki, Helsinki, Finland
   }

   \date{last revised \today}


  \abstract
  {Large area catalogs of galaxy clusters constructed from ROSAT All Sky Survey provide the base for our knowledge on the population of clusters thanks to the long-term multiwavelength efforts on their follow-up. }
  {Advent of large area photometric surveys superseding in depth previous all-sky data allows us to revisit the construction of X-ray cluster catalogs, extending the study to lower cluster masses and to higher redshifts and to provide the modelling of the selection function. }
  {We perform a wavelet detection of X-ray sources and make extensive
    simulations of the detection of clusters in the RASS data.  We
    assign an optical richness to each of the 24,788 detected X-ray
    sources in the 10,382 square degrees of SDSS BOSS area, using
    redMaPPer version 5.2. We name this survey  COnstrain Dark Energy with X-ray (CODEX) clusters. }
  { We show that there is no obvious separation of sources on galaxy clusters and AGN, based on distribution of systems on their richness. This is a combination of increasing number of galaxy groups and their selection as identification of an X-ray sources either by chance or due to groups hosting an AGN. To clean the sample, we use a cut on the optical richness at the level corresponding to the 10\% completeness of the survey and include it into the modelling of cluster selection function. We present the X-ray catalog extending to a redshift of 0.6.}
{CODEX is the first large area X-ray selected catalog of Northern clusters reaching the fluxes of $10^{-13}$ ergs s$^{-1}$ cm$^{-2}$. We provide the modelling of the sample selection and discuss  the redshift evolution of the high end of the X-ray luminosity function (XLF). Our results on $z<0.3$ XLF are in agreement with previous studies, while we provide new constraints on the $0.3<z<0.6$ XLF. We find a lack of strong redshift evolution of the XLF,  provide exact modeling of the effect of low number statistics and AGN contamination  and present the resulting constraints on the  flat $\Lambda$CDM.}


   \maketitle
%

\section{Introduction}

Many X-ray galaxy cluster catalogs rely on
identification of X-ray sources found in the ROSAT All Sky Survey as galaxy clusters \citep[see][for a summary of X-ray cluster catalogs]{piffaretti}. Given that those catalogs have been published a while ago and that they contain the brightest objects, most of the follow-up campaigns have concentrated on those clusters. In
particular, the cluster weak lensing calibration for all currently
published cosmological surveys are based on these samples. At the
moment, a difference in the weak lensing calibration of cluster masses between redshifts
below 0.3 and above have been revealed \citep{smith16}, and the importance of the selection effects at $z>0.3$ has been demonstrated \citep{kettula15}. Thus, it is important to revisit the details of the cluster selection.  

The abundance of galaxy clusters is a sensitive cosmological probe, and currently the focus of the research is to understand whether there is a tension in the reported constraints on the parameters of the $\Lambda$CDM model between clusters and CMB \citep{p16clcosmo}. It is therefore of primary importance to inspect the construction of the cluster sample and its modelling. In doing so, we will consider recently reported covariance of X-ray and optical properties \citep{farahi}, and a covariance of scatter in X-ray luminosity and shape of the cluster X-ray surface brightness \citep{kaefer19}.

\begin{figure*}

\adjustbox{trim={.0\width} {.0\height} {0.28\width} {.7\height},clip}
{\includegraphics[width=25cm]{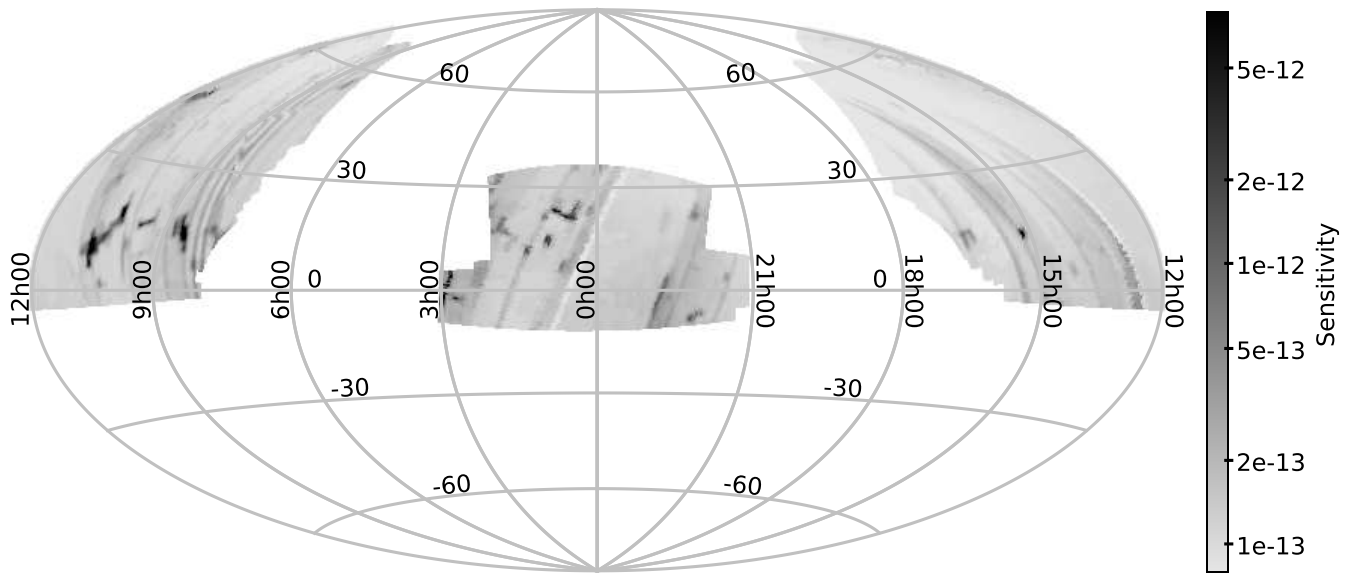}}
\caption{Aitoff projection of the sensitivity of the ROSAT All-Sky Survey data within BOSS footprint. Nominal sensitivity in the 0.5--2 keV band towards 4 counts is plotted. The units are ergs s$^{-1}$ cm$^{-2}$. The grid shows Equinox J2000.0 Equatorial coordinates. \label{sens-sky}}
\end{figure*}

With the advent of large area surveys, starting with Sloan Digital sky survey \citep[SDSS,][]{York00}, we can 
characterize the cluster identification in a fully controlled way. At the
same time, it enables an identification of a much larger sample of
X-ray sources as galaxy clusters. We present here a systematic study
of the characteristics of sources identified in this way, which we coin
as COnstrain Dark Energy with X-ray (CODEX) cluster survey.

This paper is structured as follows: in \S\ref{data} we describe the
X-ray analysis and  identification of
clusters; in \S\ref{modeling} we introduce the modelling of cluster detection based on the properties of the intracluster medium (ICM); in \S\ref{association} we present an association of detected sources with an optical galaxy cluster;  in \S\ref{exlf} we compare the observed and predicted cluster counts
based on the cosmological model and conclude in \S\ref{conclusions}.\footnote {Unless explicitly noted otherwise, all observed values quoted
  throughout this paper, are calculated adopting a $\Lambda$CDM
  cosmological model, with $H_0=70$ km s$^{-1}$ Mpc$^{-1}$,
  $\Omega_M=0.3$, $\Omega_\Lambda = 0.7$. We quote X-ray flux
  in the observer's 0.5--2.0 keV band and rest-frame luminosity in the 0.1--2.4
  keV band and provide the confidence intervals on the 68\% level. FK5 Epoch J2000.0
  coordinates are used throughout.}

\section{Data and analysis technique}
\label{data}

\subsection{RASS catalogs}

The ROSAT all-sky survey (RASS) has been an enormous legacy for X-ray
astronomy \citep[see][for a review]{Truemper93}. The whole sky has been surveyed
to an average depth of 400 seconds, yielding a total of 100,000
sources at its faint limit \citep{Voges99,Boller17}. 
Exploration of RASS sources for the purpose of identification of galaxy clusters has been primarily concentrated on the bright sub-sample \citep{b13,b17}. The main purpose of CODEX is to extend the source catalog down to the lowest fluxes accessible to RASS, reaching $10^{-13}$ ergs s$^{-1}$ cm$^{-2}$. This requires an in-depth understanding of source detection and characterization. We therefore carry out the source detection ourselves and accompany it with a detailed modeling.

RASS data is available in the form of sky images in several bands,
background images and exposure maps. We use those of Data Release
3\footnote{http://www.xray.mpe.mpg.de/rosat/survey/rass-3/main/help.html},
which contain only the photons with reliable attitude restoration \citep[for more details see][]{Boller17}.
The DR3 data consists of count maps covering an area of $\sim41$ square degree each, having some overlap between the tiles. For the source detection we use the wavelet decomposition method of  \cite{vikhl98}. We run several scales of wavelet decomposition,
starting from 2 pixels, which corresponds to 1.5$^\prime$ and extending the
search for the X-ray emission to the scales of 12$^\prime$. Larger spatial scales are important only for nearby ($z<0.1$) clusters, and within CODEX are only used for the flux refinements. Even use of the adopted scales at RASS depths tend to connect several sources \citep[see e.g.][]{mirkazemi15} and in order not to miss sources, we identify the small scales separately from the large scales and latter merge the identifications. The catalog using all scales is called C1 and the catalog derived using small scales is called C2. Given a large fraction of duplicates between the two catalogs,
we merge the sources with the offsets below $3^\prime$. Since the naming convention for the merged sources is different among
sub-projects, we present two IDs for each source, one is ``CODEX'' ID
and the other as ``SPIDERS'' ID, where SPIDERS is the spectroscopic
program on SDSS-IV \citep{dawson16,  blanton17}, which is the main source of CODEX spectroscopic identification,
with the initial results presented in \cite{clerc16} and the final catalog will be released as a part of SDSS-IV DR16 \citep[][Kirkpatrick et al., in prep.]{ahumada19, clerc20}.  

In running the wavelet detection, we set the threshold for the source detection to $4\sigma$, which shall be understood as a rejection of a possible detection of a background fluctuation at $4\sigma$ at each given place. Given 1 million independent elements in our sky reconstruction within SDSS area, one expects 53 fake sources, or a 0.998 clean catalog. However, 85\% of the sources are not clusters \citep{hasinger96}, so such a catalog as an input for galaxy cluster identification has only 15\% purity. In addition to source detection, the wavelet algorithm can be used to set the size of the region for the flux extraction by determining the region where source flux is still significantly detected. We have set this threshold to $1.6\sigma$ and in the catalog we provide the corresponding aperture in which this flux has been estimated. Our threshold is comparable to the $2\sigma$ threshold on the flux estimates, used by \cite{b04}. In restoring the full flux of the cluster we account for this aperture and in case it is comparable to the RASS PSF \citep[$\sim2^\prime$,][]{boese}, we account also for the flux lost in the wings of the PSF.   

\begin{figure}

\adjustbox{trim={.0\width} {.2\height} {0.05\width} {.1\height},clip}{\includegraphics[width=9cm]{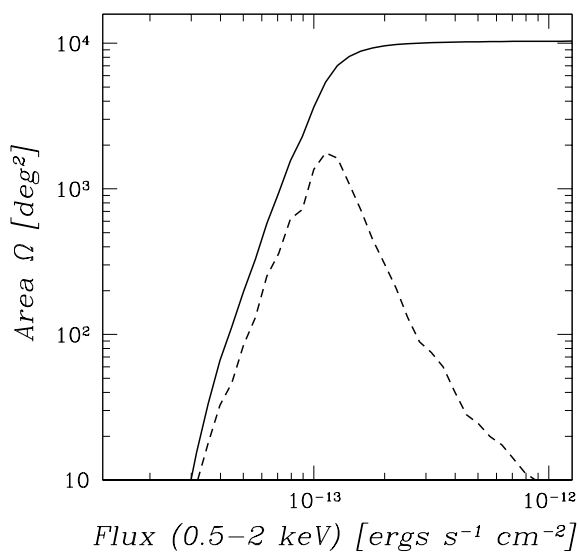}}
\caption{Cumulative ($\Omega(>S)$, solid curve) and differential ($\frac{\de \Omega}{20\de S}$, dashed curve, using $\Delta S=0.05$ dex bins in the flux) survey area as a function of flux, based on the sensitivity of the ROSAT All-Sky Survey data within BOSS footprint. Nominal sensitivity in the 0.5--2 keV band towards 4 counts is plotted.  \label{area-curve}}
\end{figure}

In the catalog, released with this paper, we remove a handful of sources (406 from a total of 90236 sources detected all sky), where detection was associated with an artefact on the image. Our flux measurements are based on a few counts, down to 4 counts (20\% of the total number of sources). The median of the count distribution is 7 counts. We use model predictions to account for the associated statistical uncertainty of the reconstruction of the X-ray Luminosity Function (XLF). 
We report the fluxes corrected for Galactic absorption. In performing this calculation, we assume a constant spectral shape of the source, and perform a correction for this assumption as a part of K-correction, which accounts for the effect of the source spectral shape (defined by temperature of the emission and the redshift). Requirements on the extragalactic sky adopted in BOSS \citep[][Baryonic Oscillations Spectroscopic Survey]{dawson13} are higher than the typical limitations considered in construction of X-ray extragalactic surveys and consequently, the variation of nH correction \citep[computed using the data from][]{kalberla}  within the survey area is small. The largest deviations in the sensitivity of the survey are driven by the variations in the exposure map of the RASS survey, which we illustrate in Fig.\ref{sens-sky}. In Fig.\ref{area-curve} we show the histogram of the survey area as a function of sensitivity.

\subsection{Source identification}

We run the redMaPPer version 5.2 \citep{rykoff14} on the position
of every source (24788 sources in the BOSS footprint), identifying the maximum richness red sequence between
redshifts 0.05 and 0.8. We search for the best optical center within 400 kpc from the X-ray center. We report the cluster richness both at X-ray and the optical positions, required corrections for the masked area of SDSS and photometric depths, which affect the error calculation for the richness. We calculate the probability of the center to be correct \citep{rykoff14}, which is useful for the weak lensing modelling \citep[e.g.][]{cibirka}. We fold the BOSS area mask into the selection of CODEX. The catalog of cluster member galaxies has been released as a target catalog of SPIDERS \citep{clerc16} and can be found online\footnote{https://www.sdss.org/dr14/data\_access/value-added-catalogs/?vac\_id=spiders-target-selection-catalogues
}.

\begin{figure}

\adjustbox{trim={.0\width} {.2\height} {0.05\width} {.1\height},clip}{\includegraphics[width=9cm]{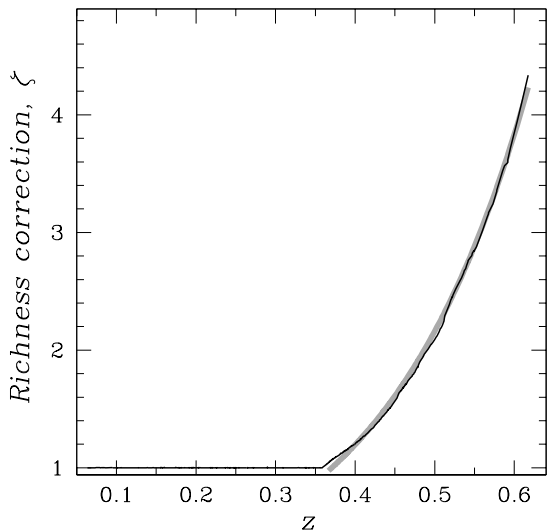}}
\caption{Multiplicative richness correction due to photometric depths of the SDSS survey. Black histogram shows the actual correction applied and grey curve shows our approximation of it as $\zeta=e^{5.5(z-0.35)}-0.12$ at $z>0.37$.   \label{zeta}}
\end{figure}

We have completed the spectroscopic follow-up campaign of CODEX clusters down to a richness of 10 through a number of
programs on SDSS-II, III and IV, as well as using
the Nordic Optical Telescope. The first results are presented in \cite{clerc16, clerc20} and Kirkpatrick et al. (in prep.), and include a full characterization of the uncertainty in the photometric redshift estimate. We report on the 100\% success rate in identification of clusters at $z<0.3$, which required 5 spectroscopic members to achieve. At higher redshift the depths of the follow-up drive the identification success and success of it reaches 100\% once we are able to target $>7$ member galaxies. The same galaxies form the bulk of the estimate of the cluster richness. Fig.\ref{zeta} shows the photometric depth correction factor used in calculating the reported SDSS cluster richness, $\zeta$, which has an exponential increase towards high-z. This is a ratio between the richness of the cluster and the actually observed part of it. High values of $\zeta$ imply that only the tip of the cluster galaxy luminosity function is observed. It also serves a description for spectroscopically confirmed sub-sample, which is a combination of the threshold for spectroscopic cluster confirmation and success rate of cluster member targeting. As the robustness of photometric identification relies on the actual number of galaxies used, consistently with other redMaPPer catalogs, we have chosen to use at least 10 member galaxies richness limit with redshift. While we also correct the richness for the masked area, the fraction of clusters with masking correction exceeding 20\% is 3\% of the sample and therefore does not require additional modelling, apart from the tests performed for galaxy cluster clustering (Lindholm et al. in prep.).

\begin{figure}

\adjustbox{trim={.0\width} {.2\height} {0.05\width} {.1\height},clip}{\includegraphics[width=9cm]{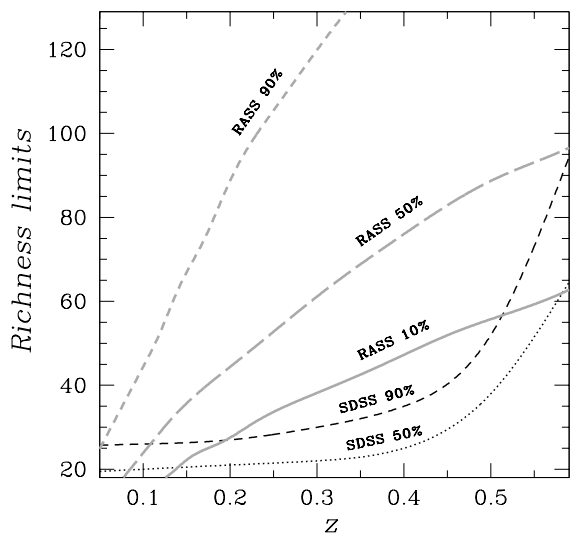}}
\caption{Richness limits of the survey. Black curves show the 90\% (dashed) and 50\% (dotted) completeness limits $P^\mathrm{SDSS}$ of redMaPPer cluster confirmation using SDSS data \citep{rykoff14}. Grey curves are the 10\% (solid), 50\% (long-dashed) and 90\% (short-dashed) completeness limits of the RASS. The 10\% curve serves also a limit for low \citep[5\%][]{klein19} contamination subsample and is adopted as our selection $P^\mathrm{RASS}$.  \label{lamlim}}
\end{figure}

The resulting redshift range of CODEX clusters is 0.05-0.65. Below a redshift of 0.1 performance of the redMaPPer has not been calibrated, and systematic offset between the photometric and spectroscopic redshifts is found \citep{clerc16}.  Large projection effects and large size of X-ray sources also require additional care. We will therefore not discuss the properties of the $z<0.1$ part of the catalog. A comparison of literature redshifts and redMaPPer measurements is also discussed in \cite{rozo15}. 

Using positions of random sources, we have estimated  a probability of the chance identification of a richness 20 source to be 10\% in each 0.1 width bin of redshift in the range $0.1<z<0.6$. In Fig.\ref{lamlim} we compare the completeness limits of the RASS and SDSS surveys towards detection of a galaxy cluster. We used the scaling relation of \cite{capasso19} to express the survey mass limits in terms of (true) richness. 
Identification of RASS sources using the  DES survey has been considered in \cite{klein19}. While this survey covers a different area, the strategy for source identification is similar. 
The 10\% sensitivity curve of the CODEX survey matches well the definition of low (5\%) contamination subsample in \cite{klein19}, which we verify using an overlap area of two surveys, located in the Stripe 82 \citep{capasso20}. In order to reduce the effect of contamination down to 5\%, we need to remove the sources with richness below the curve, and propagate this selection into the modelling. The analytical form of the selection reads:

\begin{equation}
\mathrm{exp}(\lambda_{5\%\; \mathrm{cont}}) > 22 \left( \frac{z}{0.15} \right)^{0.8}
\label{eq.prass}
\end{equation}

In addition to X-ray completeness, in Fig.\ref{lamlim} we consider the effect of optical completeness and find it to be  important for the modelling of CODEX both at $z<0.2$ and $z>0.5$. We use the following analytical function of the 50\% optical completeness ($\lambda$ denotes a natural logarithm of richness to simplify the notation for the log-normal distribution):
\begin{equation}
\lambda_{50\%}(z) = \ln (17.2 + e^{\left( \frac{z}{0.32} \right)^{2}})
\end{equation}
obtained using the tabulations of \cite{rykoff14}. We use an error function with the mean of $\lambda_{50\%}(z)$ and a $\sigma=0.2$, which reproduces the 75\% and 90\% quantiles of the distribution tabulated in  \cite{rykoff14}. We use the probability of the optical detection of the cluster in SDSS data as
\begin{equation}
P^\mathrm{SDSS}(I|\lambda,z)=1-0.5\mathrm{erfc}\left(\frac{\lambda-\lambda_{50\%}}{0.2\sqrt{2}}\right)
\label{eq.psdss}
\end{equation}
, which is discussed further in \S\ref{modeling}.

\subsection{CODEX catalog}

Once an X-ray source has an optical counterpart, we can assign a redshift to it. This allows us to compute the source rest-frame properties, such as luminosity. We apply the procedure of \cite{finoguenov07} to iteratively restore the X-ray luminosity. We obtain an initial guess on cluster mass, using $M-L_X$ relation and compute the missing source flux correction ($A$), taking into account the flux extraction aperture and the expected surface brightness profile of the source, given the mass \citep{kaefer19}. In performing mass and temperature estimates, we use the XXL $M-T$ \citep{lieu} and $L_X-T$ \citep{giles} relations, which is also consistent with CODEX weak lensing calibration of \cite{kettula15}. For small apertures, we use the PSF correction. Applying these corrections we obtain a new estimate of luminosity. We iterate this procedure 100 times. 
The resulting catalog of cluster properties is presented in Tab.1\footnote{The table will be published electronically by CDS and is temporarily available at ftp://ftp.mpe.mpg.de/people/alexis/CODEX\_aanda\_full\_flag.fits . Spectroscopic properties of CODEX clusters are released as a part of SDSS-IV DR16 under SPIDERS cluster catalog.}, with column
(1) listing CODEX source ID, (2) frequently used SPIDERS ID,  columns 3-4 providing the coordinates of the
X-ray center, column (5) providing the redMaPPer redshift, column (6-7)
providing the richness estimate and its error, column (8-9) provides the coordinates of the best optical
center, column (10) gives the
probability of this center to be correct. 
Column (11-12) lists the X-ray luminosity in the rest-frame 0.1--2.4 keV and its error, col (13) lists the temperature used in estimating the K-correction, (14-16) list $A$, $K$, PSF corrections, (17) lists the correction used in the computation of the optical richness, (18-19) the aperture-corrected flux of the source in the 0.5--2 keV band and its error, (20) flag (0 - clean catalog using Eq.\ref{eq.prass}, 1  - full catalog).  

Calculation of $L_X$ at low statistics produces on average higher values
of $L_X$, compared to the true values, due to asymmetrical shape of the Poisson distribution. The
correction for the bias depends on the mass function of clusters and
thus cosmology, in addition to the statistics of cluster detection. In \S\ref{exlf} we reproduce the resulting $L_X$ distribution, while in Tab.1 we present uncorrected properties. 

Further insights on the $L_X$ computation are available using spectroscopic identification of the sample, with the final results of the SDSS program presented in \citep{clerc20} and Kirkpatrick et al. (in prep.).

The clusters with richness greater than 60 in the previously poorly studied redshift range of
$0.4<z<0.6$ have been a subject of the CFHT follow-up \citep[e.g.][; Kiiveri et al. subm. ]{cibirka}. The SPIDERS program has enabled a dynamical mass calibration of the sample \citep{capasso19} in the full range of richness and redshifts.  Stacked weak lensing analysis of DECaLS data on $z<0.2$ CODEX clusters confirmed the results of dynamical modelling \citep{phriksee}. The results of these calibrations allow us to refine the modelling  presented in this paper and evaluate the effect of calibration on the precision of cosmological parameters.

\section{Modelling of X-ray cluster detection}
\label{modeling}

We  model the source detection as a probability given the number of detected source counts ($\eta\ob$) and the shape parameters of the surface brightness ($S_B$) distribution of the source $P(I | \eta\ob, S_B)$, where $I$ denotes the selection.  

\begin{figure}
\adjustbox{trim={.0\width} {.2\height} {0.05\width} {.1\height},clip}{
\includegraphics[width=9cm]{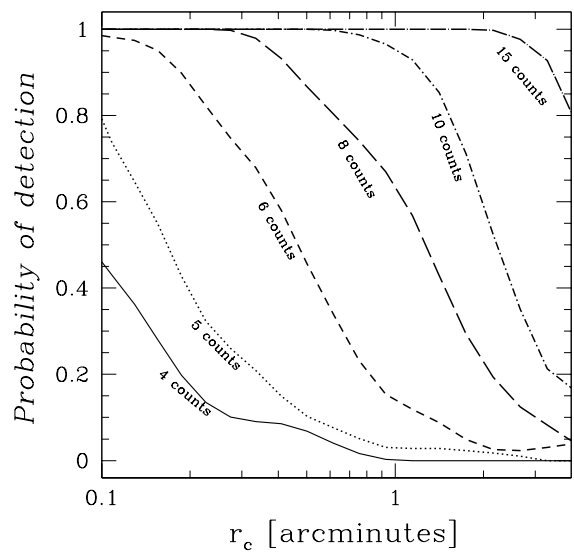}}
\caption{Probability of cluster detection as a function of its core radius $P(I|r_c,\eta\ob,\beta(\mu))$. The solid, dotted, dashed, long-dashed, dashed-dotted, long-dashed-dotted curves denote the calculation for $\eta\ob$ of 4, 5, 6, 8, 10, 15 counts, respectively. The range of core radii shown represent the cluster sample at a redshift of 0.25.  \label{det-sens}}
\end{figure}

We measure the background level in the 0.5--2 keV RASS 
images, using source-free zones. Typically, the background level corresponds to a level of { one background count in the smallest zone of detection with upto 16 background counts in the largest scales of detection}. In the model we assume a dominant contribution of Cosmic X-ray Background to the background counts and use the exposure maps as a model for its spatial distribution. Note that strong flares in RASS data release 3 have been filtered out by removing the corresponding time intervals. 

We simulate the cluster detection as a function of total number of detected cluster counts, performing a simulation of each value of the
cluster shape parameter grid 1000 times and trying 1000 realizations of background for each simulated cluster image. The grid of the cluster shape parameters samples the parameters of the $\beta$-profile of clusters, with surface brightness distributed with radius r as
\begin{equation}
S_B(r)=\left(1+\left(\frac{r}{R_c}\right)^2\right)^{-3\beta+0.5}
\end{equation}
, normalized to the total count, $\eta\true$, and sampling the distribution of core radii for clusters of given T, derived using a fixed $\beta-T$ relation \citep{kaefer19}.
We sample mass range in the 13.5-15.3 in $log_{10}(M_{200c}$) to predict the shape of the cluster ($\beta$, $R_c$), and the redshift range 0.1-0.6. We perform a Poisson realization of the simulated image, and store the results based on observed count rate ($\eta\ob$) from 3 to 30. 
As we will be using multivariate log-normal distributions throughout this paper, we conveniently define the quantities $r_c\equiv\ln (R_c)$, $l\equiv\ln (L_x$), $\mu\equiv\ln (M_{200c}$), $\lambda\equiv\ln (\mathrm{Richness})$.
We denote the obtained grid as a probability of detection given the detected counts of the source ($\eta\ob$), and shape parameters $\beta$ and $r_c$: $P(I | \eta\ob , \beta, r_c)$. To use the results of \cite{kaefer19} we substitute T with $\mu$  using $M-T$ relation of \cite{kettula15}.

As has been pointed out by \cite{kaefer19}, the cluster shape is covariant with the scatter in $M-L_X$ relation, and we will use their tabulations of the multivariate Gaussian distribution, 
$P(r_c, l | \mu)$. While \cite{kaefer19} characterize the cluster population at $z<0.1$, the importance of the considered effects is reduced at $z>0.3$, where all our clusters are nearly point-like for RASS and we deem this data sufficient. The effect of the covariant change in the core radius results in an even larger detectability of the cool core clusters. Not only do they have larger $L_X$ for a given mass, their peaked shape has a better chance of being detected \citep{eckert11}.  

In addition to covariance of X-ray luminosity and shape, we include a covariance of X-ray luminosity and richness, based on the results of \cite{farahi}, leading to 
\begin{equation}
  P(l\true,r_c,\lambda | \mu,z) =  \frac{1}{ (2 \pi)^{3/2}  | {\bm \Sigma} |^{1/2}} \exp \left[ - \frac{1}{2} {\bm X}^T {\bm \Sigma}^{-1} {\bm X}\right] \,
\end{equation}
, where the vector 
\begin{equation}
\label{eqn:scl}
{\bm X} =
 \begin{pmatrix}
  l\true - \langle l | \mu, z \rangle  \\
    r_c - \langle   r_c | \mu, z\rangle \\
    \lambda -  \langle   \lambda | \mu, z\rangle
  \end{pmatrix}
  \quad 
\end{equation}  
is defined using the scaling relations of \cite{kaefer19, mulroy19,capasso19}.  The covariance matrix ${\bm \Sigma}$ reads:
  \begin{equation}
  \quad
 \begin{pmatrix}
  \sigma^2_{l|\mu} & \rho_{l r_c |\mu} \sigma_{l | \mu}\sigma_{  r_c |\mu} & \rho_{l  \lambda|\mu}\sigma_{l| \mu}\sigma_{ \lambda |\mu} \\
  \rho_{l   r_c|\mu} \sigma_{l | \mu}\sigma_{  r_c|\mu} & \sigma^2_{  r_c|\mu} & \rho_{  r_c \lambda|\mu}\sigma_{  r_c|\mu}\sigma_{  \lambda| \mu} \\
  \rho_{l  \lambda|\mu}\sigma_{l |\mu}\sigma_{  \lambda| \mu} & \rho_{  r_c \lambda|\mu}\sigma_{  r_c|\mu}\sigma_{ \lambda| \mu}  & \sigma^2_{ \lambda |\mu}\\
  \end{pmatrix}
\end{equation}

In this work we adopt the following values  $\rho_{  r_c l|\mu}=-0.3$, $\sigma_{  r_c| \mu}=0.36$ (\citep{kaefer19}, $\sigma_{  \lambda| \mu}=0.2$ \citep{capasso19,mulroy19}, $\rho_{l   \lambda|\mu}=-0.3$ \citep{farahi}, $\sigma_{l |\mu}=0.46(1-0.61z)$ \citep{mantz16}, while no measurement of $\rho_{  r_c   \lambda|\mu}$ is published and which is set to 0 in our work. 

\begin{figure}
\adjustbox{trim={.0\width} {.2\height} {0.05\width} {.1\height},clip}{\includegraphics[width=9cm]{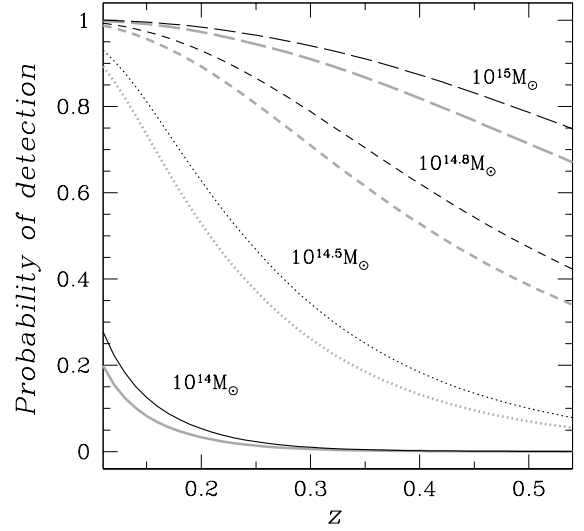}}
\caption{Probability of cluster detection as a function of redshift. The solid, dotted, dashed, long-dashed curves denote the calculation for $\mathrm{log}_{10}(M_{200c}/M_\odot)$ of 14, 14.5, 14.8, 15, respectively. Grey curves show the same calculation for clusters with richness deviating from the mass-richness relation by $+1.5\sigma_{\lambda| \mu}$.  \label{sens-mass}}
\end{figure}

The effect of richness on the selection is only by offsetting the distributions of $l$ and $r_c$,
and so for the purpose of determining the mass-richness relation, it is convenient to store just the effect of covariance on the selection function, replacing richness with  $\nu\equiv\frac{\lambda - \langle\lambda | \mu,z\rangle}{\sigma_{\lambda | \mu}}$, storing $P(I|\mu,z,\nu)$ and transforming $P(r_c, l, \lambda | \mu, z)$ to $P(r_c, l | \mu, \nu, z)P(\lambda|\mu,z)$, where only the $P(\lambda|\mu,z)$ is varied in the (external to this paper) scaling relation work. Given the freedom in treatment of the covariance $\rho_{  r_c   \lambda|\mu}$, we set it to zero, which makes $\Sigma$ a block diagonal matrix, with two $2\times2$ elements (we could also explain this by noting that $P(r_c, l | \mu, \nu, z)=P(r_c | \mu, l, z)P(l | \mu, \nu, z)$):  $\Sigma_{l\mathrm{,}\lambda}$ and $\Sigma_{l\mathrm{,}r_c}$, whose inversion is analytical, resulting in:  
\begin{multline}
   P(r_c, l | \mu, \nu, z)=\mathcal{N}\left(l-\langle l | \mu,z\rangle - {\rho_{l\nu|\mu}\nu}\sigma_{l|\mu},\sigma_{l|\mu}\sqrt{1-\rho_{l\lambda|\mu}^2}\right) \\
    \mathcal{N}\left(r_c-\langle r_c |\mu,z\rangle - \frac{\rho_{lr_c|\mu}(l-\langle l | \mu,z\rangle)\sigma_{r_c|\mu}}{\sigma_{l|\mu}\sqrt{1-\rho_{l\lambda|\mu}^2}},\sigma_{r_c|\mu}\sqrt{1-\rho_{lr_c|\mu}^2}\right) \, .
\end{multline}
The usefulness of this formula is a demonstration that covariance results in the offset of the distribution \citep[for a graphical illustration of the offset, see also][]{capasso19}.
Denoting the survey area $\Omega$ (deg$^2$) and sensitivity S (ergs s$^{-1}$ cm$^{-2}$), which includes the effects of exposure and nH, we define the survey selection function as

\begin{multline}
\label{eqn:matrix}
P(I|\mu,z,\nu) \Omega_{tot}
  = \int \de S \frac{\de \Omega}{\de S}\\
  \iiint \de l\true \de r_c \de \eta\ob P(I | \eta\ob , \beta(\mu), r_c ) P(\eta\ob | \eta\true(l\true, S, z)) \\ 
    P(r_c, l\true | \mu, \nu, z) \,
\end{multline}
, where
\begin{equation}
 P(\eta\ob | \eta\true) = \frac{(\eta\true)^{\eta\ob} e^{-\eta\true}}{ \eta\ob !} \, .
\end{equation}
Conversion of luminosity to counts uses the luminosity distance to the object $d_L(z)$, sensitivity S (counts per flux in ergs cm$^{-2}$ s$^{-1}$) and K-correction $K(\langle  T|L_X\true\rangle, z)$, and
\begin{equation}
\label{lx2eta}
\eta\true=\frac{L_X\true S}{4\pi d_L(z)^2  K(\langle  T|L_X\true\rangle, z)} \, .
\end{equation}
Fig.\ref{sens-mass} illustrates the resulting calculation for two values of $\nu$: i) $\nu=0$, i.e.
clusters following the scaling relation $\langle \lambda | \mu\rangle$ and ii) $\nu=1.5$, i.e. clusters deviating by $+1.5\sigma_{\lambda | \mu}$ from the mean relation. It demonstrates the reduced sensitivity of the survey towards clusters deviating up in richness. This matrix is used to fit the richness-mass relation (Kiiveri et al. subm.) and to constrain cosmology using the richness function (Ider Chitham et al. subm.).

The modelling of the sample takes a mass function of
clusters (for a discussion of our choice of cluster mass functions see Ider Chitham et al., subm.); predicts a covariant distribution of $L_X$ and richness for  each mass value; associates shape parameters of the cluster with
distribution of $L_X$; performs a calculation of the probability of cluster detection in each element of effective exposure and finally obtains a
corresponding effective area, which is added to the total area. 
In the above equations, $\lambda$ corresponds to the true parameter, so only the intrinsic scatter in the $\mu-  \lambda$ relation is taken into account, which for CODEX has been measured to be 0.2 \citep{capasso19, mulroy19}. In generating the SDSS richness, we need to account for the depth of the SDSS survey, using the scale value $\zeta$ (see Fig.\ref{zeta} for its redshift evolution), as discussed in \cite{capasso19}. This extra scatter is not covariant with X-ray properties, so we use $P(\lambda_\mathrm{SDSS}|\lambda,z)=\mathcal{N}\left(\lambda_\mathrm{SDSS}-\lambda,\sqrt{\frac{\zeta(z)}{\exp(\lambda)}}\right)$ 
, where we are accounting for an additional detail that the scatter in the observed richness is a function of true richness and not the mean true richness.{ We also report the measurement errors on richness, which include the effect of masking and uncertainty of local background subtraction. Comparison with deeper data reveals that account for both error terms is important.}

The expected number of clusters within a given photon count, $\Delta \eta\ob_i$, and redshift, $\Delta z_j$, bin is:
\begin{equation}
\label{eqn:NC}
 \langle N(\Delta \eta\ob_i, \Delta z_j) \rangle = \int_{\Delta z_j} \de z  \frac{\de V}{\de z \de \Omega}  (z) \\
\int \de S \frac{\de \Omega} {\de S} \int_{\Delta \eta\ob_i} \de \eta\ob \frac{\de n_\eta(\eta\ob,z)}{\de \eta\ob }
\end{equation}
, where $\Delta \Omega$ is the geometric survey area in steradians and $(\de V /\de z \de \Omega)$ is the comoving volume element, and we will be using the calculation of \cite{hogg} for the flat Universe.
The halo number density as a function of the observed photon counts $\eta\ob$ -- $(\de n(\eta\ob,S,z))/(\de \eta\ob)$ --  can be related to the theoretical halo mass function, $n(\mu,z)$, through:
\begin{multline}
\label{eqn:hmf}
\frac{\de n(\eta\ob,S,z)}{\de \eta\ob \de V}
  = 
  \iiiint \de \mu \de l\true \de r_c \de \lambda  
  P^\mathrm{RASS}(I | \lambda, z)P^\mathrm{SDSS}(I | \lambda, z) 
    \\
   P(I | \eta\ob , \beta(\mu), r_c )P(\eta\ob | \eta\true(l\true, S, z))P(r_c, l\true, \lambda | \mu,z) \frac{\de n(\mu, z) }{\de V \de\mu} \,
\end{multline}
, where 
\begin{equation}
    P^\mathrm{RASS}(I | \lambda, z)=\int_{\lambda_\mathrm{5\% cont}} \de \lambda_\mathrm{SDSS} P(\lambda_\mathrm{SDSS}|\lambda,z)
\end{equation}
corresponds to a probability of the observed richness to exceed the selection threshold needed to clean the RASS data (Eq.\ref{eq.prass}), which is a complementary error function (erfc). 
We also add an account for the optical cluster selection completeness of SDSS $P^\mathrm{SDSS}(I | \lambda, z)$, discussed above (Eq.\ref{eq.psdss}).

 Comparison with the measurements requires an additional detail, since there we  apply the aperture and PSF corrections. { When we simulate the detection, we use the full count produced by a simulated cluster, $\eta^\mathrm{ob}$. We calibrate our flux restoration routine on simulations, determining the probability of reporting $\eta^\mathrm{meas}$, given an input $\eta^\mathrm{ob}$ and the background realizations ($\eta_{bkg}(r_a)\ob -\langle\eta_{bkg}(r_a)\ob\rangle$), condition to source detection. We use Poisson probability for calculating $P(\eta_{bkg}\ob | \langle\eta_{bkg}\ob\rangle)$. We estimate $\langle\eta_{bkg}\ob\rangle$ using source-free zones in each RASS field. The function obtained this way, $P(\eta^\mathrm{meas}|\eta\ob, z, S)$, absorbs the probability of detecting  a certain extent of the source $P(r_a|r_c, \eta^\mathrm{ob}, \beta(\mu))$ and its effect on the performed extrapolation of the flux together with other assumptions made, which we repeat in our simulations: 

\begin{equation}
\eta^\mathrm{meas}=\eta^\mathrm{a} A(r_a,\langle r_c|\mu,z\rangle,\beta(\mu), R_{500})\mathrm{PSF}(r_a)\; , 
\end{equation}
where $\eta^\mathrm{meas}$ is the modelled corrected count, to be compared to that reported in the CODEX catalog, $r_a$ is the flux extraction aperture in simulations and  $\eta^\mathrm{a}$-- background corrected aperture simulated count, $R_{500}$ is the radius effectively encompassing X-ray emission \citep[e.g.][and since it is not known a priori, it requires iterations]{finoguenov07}, and is therefore effectively a function $R_{500}(\eta\ob, r_a, z, S)$. $A$ is the resulting aperture correction. To compute $R_{500}$ we use the concentration-mass relation of \cite{dm} to estimate $R_{500}$ from $M_{200}(L_X)$.
  }
The $r_a$-dependent corrections are close to unity, with typical values in the $1-1.20$ range. 
High aperture corrections, exceeding 1.5 imply that flux of the emission is much larger than its extent. These are cases of nearby sources, but also a result of flux contamination. For low-z ($z<0.15$) sources, where the aperture corrections are large, we are able to re-extract the counts using an aperture covering $R_{500}$, thus removing the need for a corrections and verify an absence of any bias in the flux estimate. { So, finally, 

\begin{multline}
\label{eqn:hmf2}
\frac{\de n(\eta^\mathrm{meas},S,z)}{\de \eta^\mathrm{meas} \de V}
  = 
  \int \de \mu \de l\true \de r_c \de \lambda \de  \eta\ob 
  P^\mathrm{RASS}(I | \lambda, z)
    \\
  P^\mathrm{SDSS}(I | \lambda, z) P(\eta^\mathrm{meas}|\eta\ob, z, S) P(r_a|r_c, \eta^\mathrm{ob}, \beta(\mu)) P(I | \eta\ob , \beta(\mu), r_c )\\
  P(\eta\ob | \eta\true(l\true, S, z))P(r_c, l\true, \lambda | \mu,z) \frac{\de n(\mu, z) }{\de V \de\mu} \, .
\end{multline}
}
It is clear that changes in the sensitivity of the survey result in the mixing of various contributions to the count distribution. It is therefore more convenient to reconstruct the observed $L_X$ distribution:
\begin{multline}
\label{eqn:matrix2}
\langle N(\Delta l_{i}\ob, \Delta z_j) \rangle = \int_{\Delta z_j} \de z \frac{\de V}{\de z \de \Omega}  (z)
\int \de S \frac{\de \Omega}{\de S}\\
\int_{\eta^\mathrm{meas}(\Delta l_{i}\ob)}\de \eta^\mathrm{meas} \frac{\de n_\eta(\eta^\mathrm{meas},S,z)}{\de \eta^\mathrm{meas} \de V}
 \, 
\end{multline}
.

{ Practical implementation of Eq. \ref{eqn:hmf2}, uses low and high statistics approximation with a switch at 20 detected counts. For low statistics, we find $\eta^\mathrm{a}\approx \eta\ob$ with a much lower scatter, compared to an attempt to reproduce $\eta^\mathrm{meas}$. So, in the modelling the counts, performed in Fig.\ref{xlf2}, we  mix aperture (dominant in high-z XLF and lowest $L_X$  low-z bin) and full  X-ray luminosities from the data and do the same in the modeling.}

\section{Modelling of the association between X-ray source and optical cluster}
\label{association}

We consider the following processes to result in the association between
an X-ray source and the optical cluster

\begin{enumerate}

\item Chance association between an X-ray source and an optical cluster.

\item Detection of the optical cluster due to AGN activity of its member galaxies.

\item Detection of the optical cluster due to the thermal emission of ICM

\end{enumerate}

The probability of chance identification has been calculated by
placing random points on the sky and running the redMaPPer algorithm on
them.  We obtain the probability of chance identification as a
function of redshift for two cuts in detected optical richness, 10 and
20. While within a factor of 1.5, there are no changes in the chance
identification rate with redshift, there is a strong increase
in the chance identification towards low values of richness, going
from 10\% for richness of 20 to 30\% for richness of 10.

\begin{figure*}
\adjustbox{trim={.0\width} {.2\height} {0.05\width} {.1\height},clip}
{\includegraphics[width=9cm]{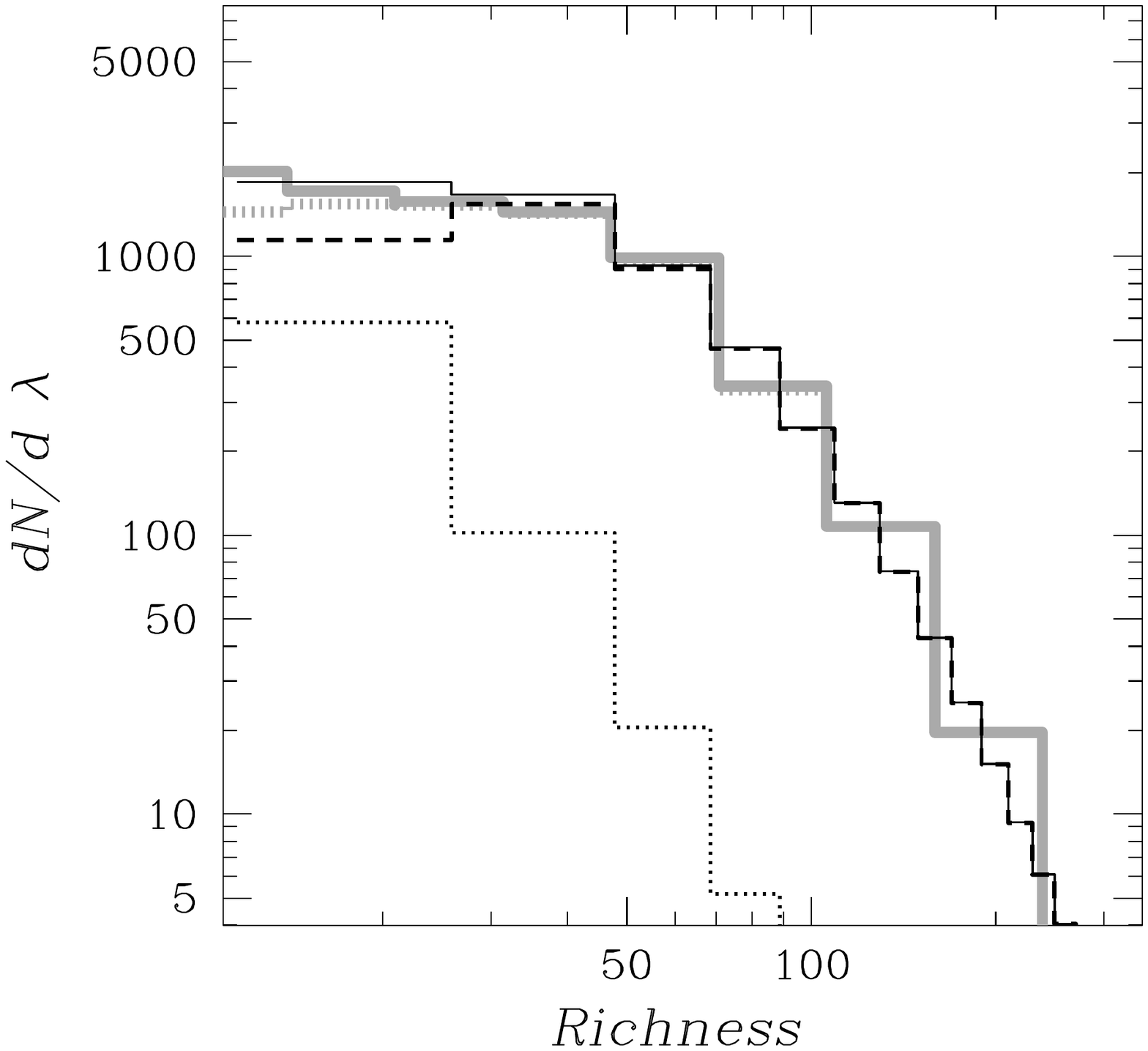}}
\adjustbox{trim={.0\width} {.2\height} {0.05\width} {.1\height},clip}
{\includegraphics[width=9cm]{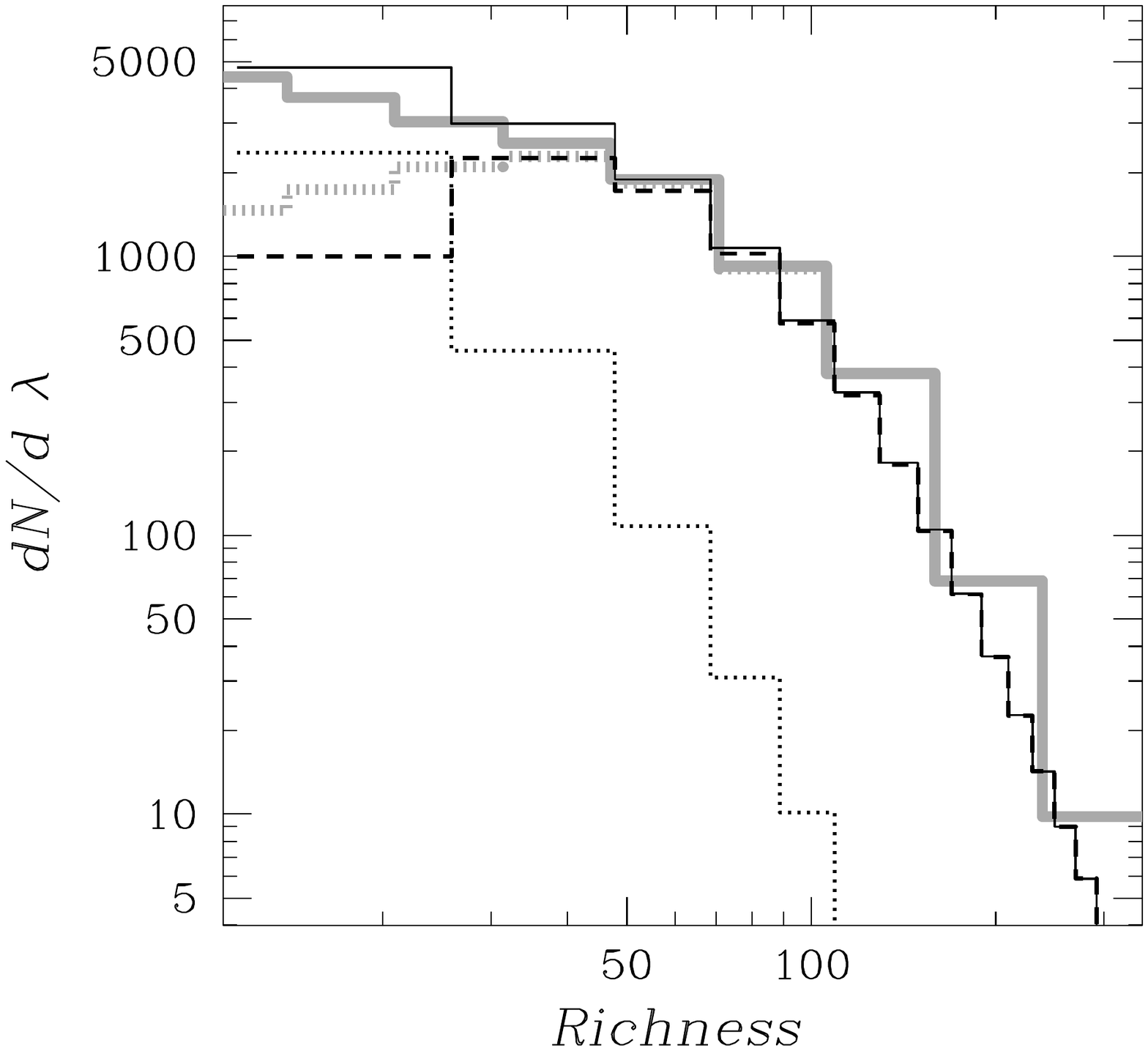}}
\caption{Richness function of CODEX clusters in the $0.1<z<0.3$ (left panel) and $0.1<z<0.6$ range (right panel). The solid grey histogram shows the data, the dotted histogram shows the contribution to the total counts from the clusters detected through their AGN activity, the dashed histogram show the contribution from clusters detected through their thermal emission and the solid histogram show the total expected number of detection, which provides a good match to the data. 
The dotted grey histogram shows the data with excision of points deviating beyond the 2$\sigma$ from the richness$-L_X$ relation. 
\label{clf}}
\end{figure*}

The probability of the identification of a cluster predominantly through its
AGN activity is driven by the probability of a cluster to host an AGN,
which is given by the AGN halo occupation distribution (HOD), and a probability of AGN to have a
certain luminosity, which is given by the AGN X-ray luminosity
function (AGN XLF). We use the HOD results of \cite{allevato12} and the \cite{ebrero} luminosity function for 0.5--2 keV to perform the
calculation. The typical luminosity of AGN, calculated using the AGN XLF,
is $10^{44}$ ergs s$^{-1}$, with the probability of finding such an
AGN in a cluster of $0.05 \times (1+z)^{3.3}$. There is no dependence on halo mass at $M_{200c}>10^{13}M_\odot$, predicted by the model.  This modelling allows us to conclude  that AGNs only provide a modest contamination to cluster luminosity, important only at lower redshifts, where our sensitivity is below the typical AGN luminosity.  According to this modelling, the main
contribution to cluster counts are AGNs detected in galaxy
groups. High X-ray luminosity and low optical richness systems are
therefore regarded as AGNs in groups or chance identification.

In Fig.\ref{clf} we compare the measured cluster richness function
with the prediction based on $\Lambda$CDM and our AGN contamination model. AGN
luminosity produces an additional component which at zero order is
simply a fraction of all clusters of a given richness that we have not yet
detected. The evolution of the fraction of the detected clusters as a
function of redshift is due to two competing effects, evolution of the
AGN XLF, and evolution of the threshold luminosity of AGN,
which leads to a decreased AGN detectivity per
cluster.

In addition to detection of new systems, AGNs can contribute to the
total luminosity of the clusters, selected primarily by the ICM
luminosity. This contribution is discussed in \cite{clerc16} and is below the 10\% level.

So, how do these results compare with contamination calculation of \cite{klein19}? As we have mentioned the contaminated zone outlined in \cite{klein19} corresponds to the 10\% X-ray completeness curve in our calculation. There, truly detected clusters are 10\%, while the rest 90\% can be identified by chance (richness-dependent process), or by AGN activity (nearly richness independent). As we mentioned the AGN activity yields 2\% identification, chance identification is at most 10\% for richness of 20. The fractional importance of the contamination is therefore 50\% of the total at lowest value of richness considered here, and drops to 9\% at high redshift where it is dominated by AGN HOD while chance identifications are rare as number of clusters of high richness is low. This consideration allows us to conclude that contamination is indeed driven by the lack of real detections.

\begin{figure}
\adjustbox{trim={.0\width} {.2\height} {0.05\width} {.1\height},clip}
{\includegraphics[width=9cm]{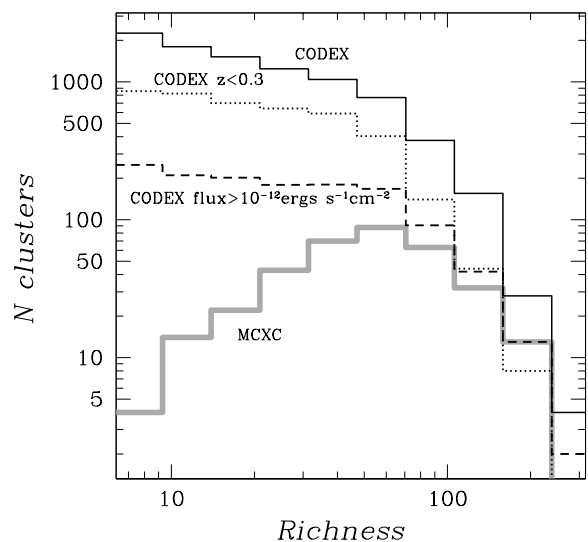}}
\caption{Richness distribution of CODEX clusters (solid black
  histogram), compared to a subsample of CODEX clusters with flux
  above $10^{-12}$ ergs cm$^{-2}$ s$^{-1}$ (dashed black histogram),
  $z<0.3$ CODEX clusters (dotted black histogram) and the matched MCXC
  clusters (solid grey histogram). The comparison shows a deficiency
  in the low richness identifications present in the MCXC
  catalog. \label{r_com}}
\end{figure}

In Fig.\ref{r_com} we compare the richness distributions of CODEX
clusters, its subsamples, based on flux and redshift and the
literature sample, MCXC \citep{piffaretti}, matched to CODEX
clusters in order to obtain a richness estimate. The literature sample
is primarily composed of the bright X-ray clusters and its distribution does not
significantly change by imposing a cut on the flux of $10^{-12}$ ergs
cm$^{-2}$ s$^{-1}$ . We imposed a similar flux cut to the CODEX sample
in order to illustrate the effect of a different flux. We also test a
redshift cut of $z<0.3$ on the CODEX sample to eliminate the effect
of the noise in the optical data. The comparison points out that the
literature sample of clusters systematically lacks identification of
clusters of richness below 70 (obtained by restricting the comparison to
both low z and high flux subsample), which corresponds to the clusters
at the limit of the Abell cluster definition. Some of the literature \citep[e.g.][]{b02} removal of sources based on the
identification with an optical QSO, which can partially account for
the difference. Also, NORAS survey that entered MCXC was not complete \citep{b17}. 

{ In the parallel effort of \cite{comparat19}, RASS sources are identified using NWAY \citep{salvato18}. Using the overlap of CODEX catalog with the DR16 area, we have verified our estimates of contamination based on the alternative matches, provided by NWAY. Using the 1281 clusters in the overlap clean catalog, we find 78 sources identified as AGN, 4 as BL LAC and 3 as stars, with 22 sources having matching redshifts to CODEX clusters within the redMaPPer errors. This corresponds to a 4.9\% chance association and a 1.7\% chance of flux contamination. Similarly, for a total sample of 6240 CODEX clusters in the DR16 area, 975 (263 matching in redshift) are identified as AGN, 26 (6 matching in redshift) as BL LACs and 26 as stars. This corresponds to a 12.1\% chance association and a 4.3\% chance of flux contamination for the full sample. These numbers correspond well to our estimates of catalog purity and illustrate the effectiveness of our catalog cleaning method. In \cite{clerc20} we have estimated the effect of AGN contamination on high-z XLF as 15\%, while it is about 2\% effect for the low-z XLF. In the modelling, we correct the high-z XLF for this and add 5\% fractional errors to account for uncertainty of the correction. We attribute the increased importance of AGN in high-z part of the sample to the increased scatter in SDSS richness estimate, which reduces the efficiency of cleaning the sample at high-z. For improvements using deeper photometry we refer the reader to Ider Chitham et al. (subm.).}

\section{Evolution of cluster X-ray Luminosity Function}
\label{exlf}

One of the direct measurements that CODEX provides is the evolution of
the XLF of galaxy clusters. The evolution of XLF combines a decrease in the number of clusters of given
mass with higher X-ray luminosity per given mass at higher redshifts. 
While, we are using red sequence redshifts in calculation of the X-ray luminosity, we have verified that use of spectroscopic redshift does not change the XLF \citep{clerc20} and so we can omit the integration over the redshift uncertainties. 

\begin{figure}
\adjustbox{trim={.0\width} {.2\height} {0.05\width} {.1\height},clip}{\includegraphics[width=9cm]{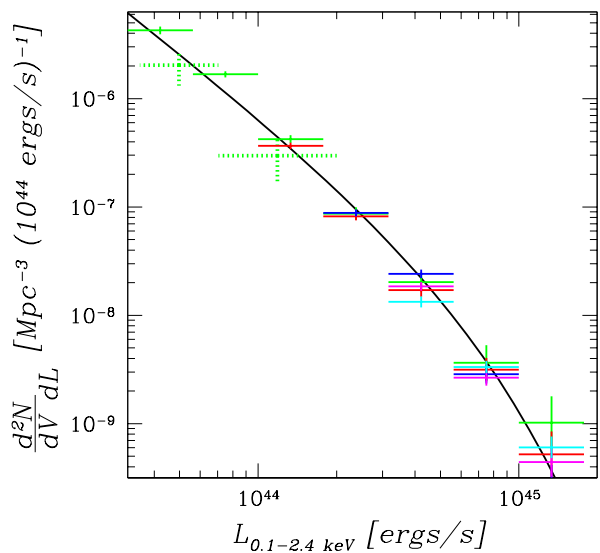}}
\caption{X-ray luminosity function of CODEX clusters in the redshift range 0.1--0.6 (0.1--0.2 green, 0.2--0.3 red, 0.3--0.4 blue, 0.4--0.5 magenta, 0.5--0.6 cyan). At low $L_X$ we also show the COSMOS XLF, for comparison (the green dotted crosses). The solid black curve shows a Schechter function description of REFLEX XLF \citep{b02}. Our results agree well with the REFLEX and reveal no strong redshift evolution of XLF. \label{xlf}}
\end{figure}

In Fig.\ref{xlf} we present CODEX constraints on the evolution of the cluster XLF measured in the redshift interval $0.1<z<0.6$ using bins of redshift of 0.1. We only show the data without strong (exceeding a factor of 2) completeness correction, as those are sensitive to both the adopted scaling relations and impact of the assumed distance-redshift relation on detection statistics.  The completeness correction is calculated by rationing the predicted distribution of clusters on luminosity accounting for the sample properties, described above (Eqs.\ref{eqn:hmf}-\ref{eqn:matrix2}) and the one assuming no selection effects and infinite statistics: 
\begin{multline}
\langle N(\Delta l_{i}\true, \Delta z_j) \rangle = \int_{\Delta z_j}\de z \frac{\de V(\Omega )}{\de z } \int_{\Delta l_i} \de l\true \\
\int \de\mu 
P(l\true | \mu,z) \frac{\de n(\mu, z) }{\de V \de\mu}
\end{multline}
The main results seen in Fig.\ref{xlf} are i) an agreement with the XLF determined from other low-z ($z<0.3$) studies, and ii) a lack of strong evolution of the XLF with redshift.

\begin{figure}
\adjustbox{trim={.0\width} {.2\height} {0.05\width} {.1\height},clip}{\includegraphics[width=9cm]{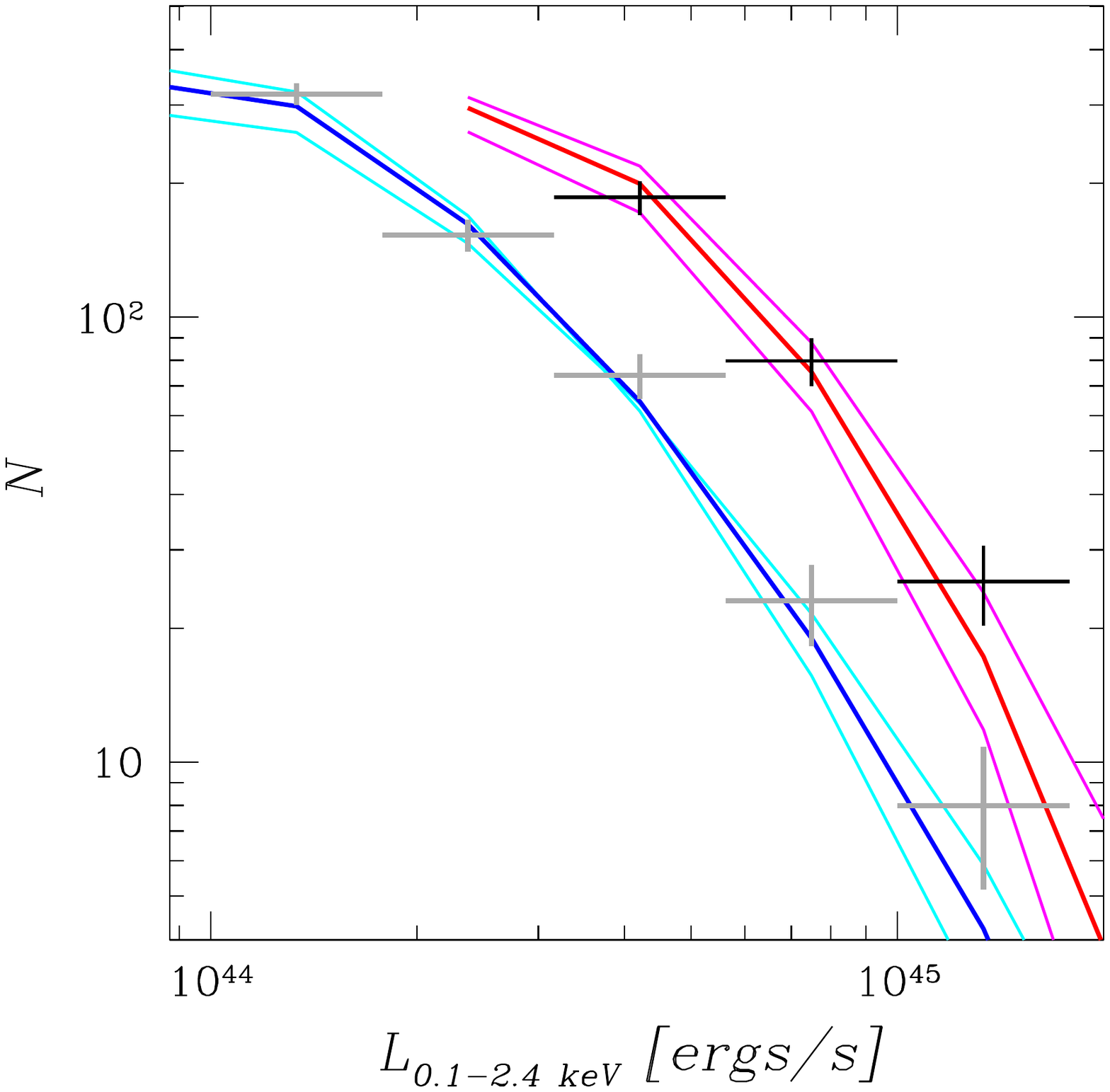}}
\caption{Number of CODEX clusters in the redshift range 0.1--0.3 (grey) and 0.3--0.6 (black), compared to the allowed range of $\Lambda$CDM cosmological parameters. Low-z solution are shown in blue (best fit) and cyan (extremes of 95\% confidence interval), high-z -- in red (best-fit) and magenta (extremes of 95\% confidence interval). The numbers are corrected for the expected AGN contamination and luminosity is a mix of corrected and uncorrected values for high and low statistics subsample.  \label{xlf2}}
\end{figure}

In order to compare the observed luminosity function with the expectations of different cosmological models, we need to adopt a mass calibration. At the moment, the $0.3<z<0.6$ part of the CODEX scaling relations have not yet tightly constrained, while the available studies extending to a redshift of one argue in favor of the self-similar scaling for these relations outside of the cool core \citep{mcdonald}. So, here we will consider whether the observed lack of evolution of XLF can be explained by a combination of self-similar evolution of scaling relations and a change in some basic cosmological parameters, limiting the study to $\Omega_M$ and $\sigma_8$. The dominant systematics of this assumption is a lack of constraints on the evolution of the cool core and AGN contamination in the CODEX high-z sample. These need to be understood better by future work. 
Adjusting the mass calibration for changing the $\Omega_M$ is not required in the scaling relations based on dynamical mass estimates done with respect to the critical density, as $M_{200c}\sim \sigma^3 \sqrt{\rho_{crit}}$; assumption of the geometry of Universe in calculation of $L_X$ cancels out by the calibration procedure, which establishes the link to the mass and so does not need to be updated on average. And the mass is both measured and used in defining the mass function with the same scaling for the Hubble constant. However, since our modelling of the high-z sample considers a self-similar evolution of scaling relations, instead of direct determination, we need to allow for an effect of geometry of the Universe on $L_X$. 

\begin{figure}
\adjustbox{trim={.0\width} {.23\height} {0.0\width} {.1\height},clip}
{\includegraphics[width=9cm]{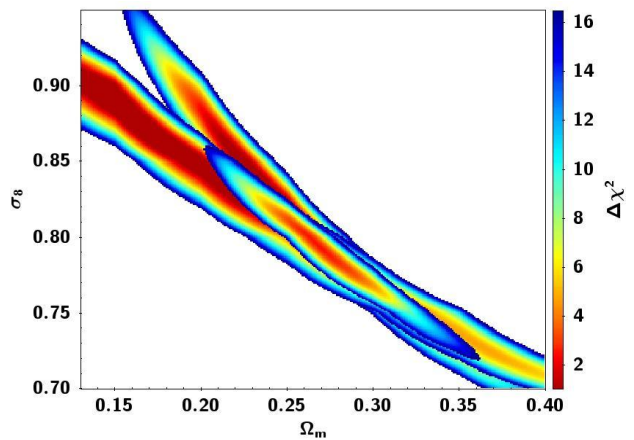}}
\caption{Constraints on the flat $\Lambda$CDM model by the CODEX cluster XLF. We show the $0.1<z<0.3$ (drawn below all other shades, lowest in the right corner) and $0.3<z<0.6$ (highest in the right corner) constraints, as well as their overlap (drawn atop of all the areas). The color shows the $\Delta\chi^2$ from the minimum  with 2.3, 4.6, 9.2 corresponding to 68, 90 and 99\% confidence level for two parameters \citep{lampton}.  \label{s8}}
\end{figure}

In Fig.\ref{xlf2} we compare the observed number of clusters with the results of the modelling presented in this paper. In Fig.\ref{s8}, we present the associated cosmological parameters. We use two redshift bins to simplify the presentation of the results:  low redshift bin $0.1<z<0.3$ and high redshift bin, $0.3<z<0.6$. We estimate the errors ($\sigma_{ij}$) based on the measured number of clusters. Given the large area of the survey and the large redshift bins used, we can ignore the sample variance term in the mass-function calculation. In Fig.\ref{xlf2} we plot the models that satisfy both high-z and low-z sample. 

Smaller values of $\Omega_M$ predict slower evolution of the mass function and larger volume, partially compensated by the slower evolution of the scaling relation, and smaller cluster X-ray detected count-rates, which would be converted into smaller $L_X$ under a fixed cosmology.
The normalization of the XLF can be adjusted by changing $\sigma_8$, but it is constrained by the slope of the XLF, which in our case is well measured only at low-z.

In calculating the best fit, we used a grid of models, covering values of $\Omega_m$ in the $(0.1-0.4)$ interval and the values of $\sigma_8$ in the $(0.7-1.0)$ interval.
We compute the likelihood of the solution using $\chi^2$. The minimum of the 
\begin{equation}
\chi^2=\sum \frac{(N_{ij}\ob - \langle N(\Delta l_{i}\ob, \Delta z_j)\rangle )^2}{\sigma_{ij}^2}
\end{equation}
is comparable with the number of degrees of freedom and thus the solutions are statistically acceptable. To compute the error intervals on cosmological parameters, we use the deviations from the minimum. We quote the errors associated with two parameters of interest, so $1\sigma$ corresponds to a $\Delta\chi^2=2.3$.  

The cosmology of low-z sample is comparable with previous similar studies \citep{b14}, with the slight differences in the best-fit values primarily due to the adopted mass calibration, well within the reported associated uncertainties. Thus, our revision in the cluster identification did not result in the change of the cosmological constraints coming from low-z RASS surveys, with a possible exception of the work of \cite{mantz16}, where we disagree on the adopted $M-L_X$ relation. Our relation has a 10\% lower normalization in mass, coming from LoCuSS and CCCP studies \citep[for a careful discussion of the problem, see][]{smith16}, and confirmed by the CODEX mass calibration efforts \citep{capasso20, phriksee}. We deem our low-z calibration to be good to 5\% in mass, which results in the associated systematical uncertainty in $\Omega_m$ of 0.015. 
For the discussion of comparison of low and high-z XLF,  this uncertainty would shift the final solution, but does not result in a larger overlap in the solutions and so, we do not show it in Fig.\ref{s8}.
The main result, seen in Fig.\ref{s8} consists in finding, { that after correction for the flux reconstruction at low statistics and AGN contamination, the cosmological constraints coming from the high-z CODEX cluster sample largely overlap with those of low redshift, leading to a large allowed range of cosmological parameters}. Within a flat $\Lambda$CDM and an assumption of self-similar evolution of scaling relations, the required cosmological parameters are $\Omega_m=0.270\pm0.06\pm0.015(syst.)$ and $\sigma_8=0.79\pm0.05\pm0.015(syst.)$,  quoting 68\% confidence interval, with a  combined $\sigma_8(\Omega_m/0.3)^{0.3}=0.77\pm0.015\pm0.015(syst.)$. 
Comparison to the literature on galaxy clusters, our solution covers the parameter space in common with all cluster surveys \citep{b14, henry09, bocquet19, vikhlinin09} and Fig.\ref{s8} can serve as a forecast for the importance of the calibration of scaling relation work at high-z.

\section{Conclusions}
\label{conclusions}

We present a new large catalog of X-ray clusters detected in the SDSS area. The catalog is constructed with the aim of efficient spectroscopic follow-up program of SDSS, which has completed data acquisition process in March 2019 and this paper describes the catalog construction to accompany the final (DR16) data release.  

Despite the low photon statistics of RASS, we show that we can convincingly model the survey selection function. We point out that low-richness clusters are underrepresented in the identification process of X-ray clusters, and this needs to be included in the modelling of the sample selection. We provide the forward modelling of such a selection and apply it to the sample to construct the  X-ray luminosity functions of the survey. 
Our main result consists in the lack of evolution of cluster XLF. { However, low statistics associated with the flux measurement at high-z redshift and AGN contamination were found to be of large importance in understanding the effect, while the associated constraints on the cosmological parameters are consistent with those of other cluster surveys.} As with most new cluster samples, more work on understanding the properties of clusters will serve towards improving the robustness of the results and uniqueness of CODEX consists in its largest calibration database on cluster dynamics, which is yet to be fully explored.

\section*{Acknowledgements}
We thank the referee for their comments which improved the quality of this paper. 
Funding for the Sloan Digital Sky Survey IV has been provided by the Alfred P. Sloan Foundation, the U.S. Department of Energy Office of Science, and the Participating Institutions. SDSS-IV acknowledges
support and resources from the Center for High-Performance Computing at
the University of Utah. The SDSS web site is www.sdss.org.

SDSS-IV is managed by the Astrophysical Research Consortium for the 
Participating Institutions of the SDSS Collaboration including the 
Brazilian Participation Group, the Carnegie Institution for Science, 
Carnegie Mellon University, the Chilean Participation Group, the French Participation Group, Harvard-Smithsonian Center for Astrophysics, 
Instituto de Astrof\'isica de Canarias, The Johns Hopkins University, Kavli Institute for the Physics and Mathematics of the Universe (IPMU) / 
University of Tokyo, the Korean Participation Group, Lawrence Berkeley National Laboratory, 
Leibniz Institut f\"ur Astrophysik Potsdam (AIP),  
Max-Planck-Institut f\"ur Astronomie (MPIA Heidelberg), 
Max-Planck-Institut f\"ur Astrophysik (MPA Garching), 
Max-Planck-Institut f\"ur Extraterrestrische Physik (MPE), 
National Astronomical Observatories of China, New Mexico State University, 
New York University, University of Notre Dame, 
Observat\'ario Nacional / MCTI, The Ohio State University, 
Pennsylvania State University, Shanghai Astronomical Observatory, 
United Kingdom Participation Group,
Universidad Nacional Aut\'onoma de M\'exico, University of Arizona, 
University of Colorado Boulder, University of Oxford, University of Portsmouth, 
University of Utah, University of Virginia, University of Washington, University of Wisconsin, 
Vanderbilt University, and Yale University.

Based on observations made with the Nordic Optical Telescope, operated by the Nordic Optical Telescope Scientific Association at the Observatorio del Roque de los Muchachos, La Palma, Spain, of the Instituto de Astrofisica de Canarias.

\end{document}